\documentclass{sig-alternate}
\usepackage{url}
\usepackage{graphicx}
\usepackage{url}
\usepackage{booktabs}
\usepackage{flushend}
\usepackage{microtype}
\usepackage{tabulary}
\usepackage{pifont}

\author{
Eduardo Graells-Garrido\thanks{Contact: \url{eduardo.graells@telefonica.com}} \\
\affaddr{Universitat Pompeu Fabra} \\
\affaddr{Barcelona, Spain} \and
Mounia Lalmas \\
\affaddr{Yahoo Labs} \\
\affaddr{London, UK} \and
Ricardo Baeza-Yates \\
\affaddr{Yahoo Labs} \\
\affaddr{Barcelona, Spain}
}

\title{Sentiment Visualisation Widgets for Exploratory Search}

\newcommand{\spara}[1]{\smallskip\noindent{\bf #1}}

\makeatletter

\toappear{Presented at the Social Personalization Workshop held jointly with ACM Hypertext 2014.}

\begin{document}

\maketitle

\begin{abstract}
This paper proposes the usage of \emph{visualisation widgets} for exploratory search with \emph{sentiment} as a facet. 
Starting from specific design goals for depiction of ambivalence in sentiment, two visualization widgets were implemented: \emph{scatter plot} and \emph{parallel coordinates}.
Those widgets were evaluated against a text baseline in a small-scale usability study with exploratory tasks using Wikipedia as dataset. 
The study results indicate that users spend more time browsing with scatter plots in a positive way. 
A post-hoc analysis of individual differences in behavior revealed that when considering two types of users, \emph{explorers} and \emph{achievers}, engagement with scatter plots is positive and significantly greater \textit{when users are explorers}.
We discuss the implications of these findings for sentiment-based exploratory search and personalised user interfaces.
\end{abstract}


\section{Introduction}
Search is a common activity on the web today, performed by almost everyone. 
Even though search  engines have been present for many years on the web, today most of them still have the initial text-based interface, which is shown to all users, in spite of the emergence of several paradigms in information seeking and user modeling that could be used to personalise it. 

One of those paradigms in information seeking is \emph{Exploratory Search} \cite{marchionini2006exploratory}, where a concrete information need is not always present and information seekers usually engage in learning and investigation strategies instead of plain lookup of documents.
One way to support exploratory search is by using faceted search interfaces \cite{hearst2009search}, where information seekers have access to several orthogonal dimensions of the information space even when there is no explicit information need. 
This approach allows information seekers to explore the information space without writing a query. 
However, its implementation requires a structure in the underlying data that is not always available.
A solution to this is to extract meta-data from the information space to provide the needed structure. 
In this paper we adopt this approach to build a facet for an unstructured information space, by using attributes annotated in text documents calculated through \emph{sentiment analysis}~\cite{pang2008opinion}. 

Users are getting used to see and understand emotional annotations in text, as popular websites such as news outlets and e-commerce sites have user ratings and reviews, which are inherently emotional. 
However, to the extent of our knowledge, this emotionality inherent in the text has not been exploited to encourage exploratory search. 
This is somewhat surprising as it is not uncommon for information seekers to have sentiment in mind when performing some tasks, for instance, when browsing user reviews to find restaurants, movies, places, or other things where the  emotional or affective responses of other users are important.
When sentiment is actually depicted in these scenarios, its depiction is usually focused on a single variable that goes from negativity to positivity, and often this variable is discrete, as in the case of a simple text classification of \emph{negative, neutral or positive}, or a \emph{n-star ratings}. 
Using only a single variable hides the richness of the various sources of sentiment and their distribution. For instance, in review sites, the only way to find the sentiment diversity is by manually browsing the list of reviews, as a \emph{n-star} rating simply displays an average.

Most sentiment depictions do not consider the ambivalence present in text, which means that a document may have both positive and negative content at the same time. 
In our approach we build \emph{visualisation widgets} \cite{dork2008visgets} where the widget visualises ambivalent sentiment as a facet for search results. 
Although this may be feasible using the typical text widgets used in faceted interfaces, our work focuses on visualisation to provide an exploratory experience that is engaging. 
In this regard, our research question is: \textbf{do visual approaches foster exploration in a sentiment-based exploratory search setting?} 
To answer this question, we defined a set of design goals for visualisation widgets in our setting. 
We fulfilled those goals with two interactive visualisations based on known paradigms: \emph{scatter plots} and \emph{parallel coordinates}, and tested these visual approaches against a text-based baseline. 
We performed quantitative and qualitative analysis to analyse the results and see if  exploration using sentiment-based visualisation widgets is fostered from a user engagement perspective.

As information space for a case study we chose Wikipedia, an open encyclopedia where anyone can contribute and edit articles. 
Wikipedia is a prominent social media platform, which contains articles with inherent sentimental content \cite{mejova2013searching}. 
In addition, its users, both readers and editors, search more on average than those never or hardly using Wikipedia \cite{west2012data}. 
This prominence of search in Wikipedia, its publicly available content and the existence of sentiment in it, made it a good candidate to use as basis to evaluate our visualisation widgets.

This paper contributes a user evaluation of exploratory behavior in the presence of sentiment in both user intent and information space. 
Based on the study results, we show that users are spend more time performing tasks when using scatter plots. 
This additional time is explained by positive engagement \textit{when users are explorers}, based on qualitative feedback and the analysis of individual differences \cite{chen2000individual}.
The analysis of individual differences was based on how users interact with search interfaces: we identified two types of users, \emph{explorers} and \emph{achievers} \cite{bartle1996hearts}. 
Our results suggest that scatter plots are more suitable for explorers, as they significantly increase engagement, opening a path to research which visualisations or interface elements are more suited for achievers, for whom we did not find a particular visualization that increased engagement.

\section{Related Work}

Although bar and pie charts are common depictions to visualise sentiment, there are other approaches to visualise it. 
In \cite{gregory2006user}, affect in document collections is visualised with \textit{wind rose charts}.
\textit{Heatmaps} are used in \cite{diakopoulos2010diamonds} to encode the average sentiment of a period of time. 
In the context sentiment in reviews, \cite{alperopinionblocks} used \textit{histograms} and \cite{carenini2006interactive} used \textit{treemaps}.
\textit{Scatter plots} are used in \cite{panger2013visualizing} to visualise ambivalence in public opinions. 
This is the most similar work to ours from a visualisation perspective, as other previous work focused on unidimensional color-coding of sentiment. 
We also use \emph{parallel coordinates} \cite{inselberg1985plane}, which have not been used before in this context to the extent of our knowledge.

\emph{We Feel Fine} \cite{kamvar2011we} is a search engine where information seekers can answer questions with an explicit sentiment component such as \emph{``how did the U.S. feel when Obama was elected?''} and obtain a visualisation of search results. 
The purpose of the visual depiction is artistic, and results can be filtered through facets of meta-data such as gender, age and mood. 
With regard to \emph{visualisation widgets}, \cite{dork2008visgets} depicts facets such as time, geo-location and topics. 
In \cite{clarkson2009resultmaps}, treemaps are used to depict a hierarchical facet. 
It was found that the usage of visualisation had positive impact on perceived task difficulty, repository understanding and enjoyment.
Our work extends \cite{dork2008visgets}, as we present widgets for a specific facet that could be used among other widgets.

In many search scenarios the information seeker is not an expert who has to perform a concrete, specialised task. 
Hence, non-experts have a diversity of expertise, knowledge and experience with computer systems.
Because not even two persons are equal, the study of \emph{individual differences} \cite{chen2000individual} proves to be useful, as it allows to find which factors, from demographic, cultural and behavioral, have impact on user modeling and user generated content.
In informational contexts, personality traits have been considered to define a user taxonomy of \emph{fast surfers, broad scanners and deep divers} \cite{heinstrom2002fast}. 
In virtual worlds, a popular taxonomy is based on how people interact with the world: \emph{achievers} and \emph{explorers} \cite{bartle1996hearts}. 
We consider the latter taxonomy as a first step towards more complex ones.

\section{Sentiment Visualisation}

We start from a scenario where the information seeker already has a query, but one that is not necessarily final.
We consider learning and investigation activities \cite{marchionini2006exploratory} as focus for design goals. Our design goals are:

\spara{Depict ambivalence.}
Typical sentiment depictions only show one sentiment attribute, often as a mixture of both positivity and negativity to find out which one is prevalent. 
However, ambivalence is present in many categories and genres of textual content, including public discourse, fiction and news articles. 
Information seekers should be able to see the duality of sentiment in text, depicted in terms of \emph{positivity} and \emph{negativity}, or ambivalence directly as in \cite{panger2013visualizing}. 

\spara{Show sentiment distribution.}
Following the scenario presented by \cite{kamvar2011we}, questions such as \emph{``How did the U.S. feel when Obama was elected?''} have an implicit request for seeing distribution and an explicit request for seeing sentiment. 

\spara{Allow sentiment filtering.}
The interface of \cite{kamvar2011we} uses sentiment keywords such as \emph{mood}, \emph{sad}, \emph{happy}, \emph{depressed}, to filter results according to emotion. 
In text query interfaces, information seekers depend on the context at hand, and a keyword search may exclude the desired sentimentality because the information seeker did not use ``matching'' keywords. 
Visual filtering would remove the burden of writing the correct keywords from users and provide a more flexible tool for filtering according to emotion.

A visualisation widget that conforms to these design goals will allow information seekers to understand how sentiment is distributed in an information space, to see the ambivalence present in it and to filter documents in order to learn and investigate according to their own criteria.

\subsection{Visualisation Widgets}

\begin{figure}[htb]
\centering
\includegraphics[width=0.48\textwidth]{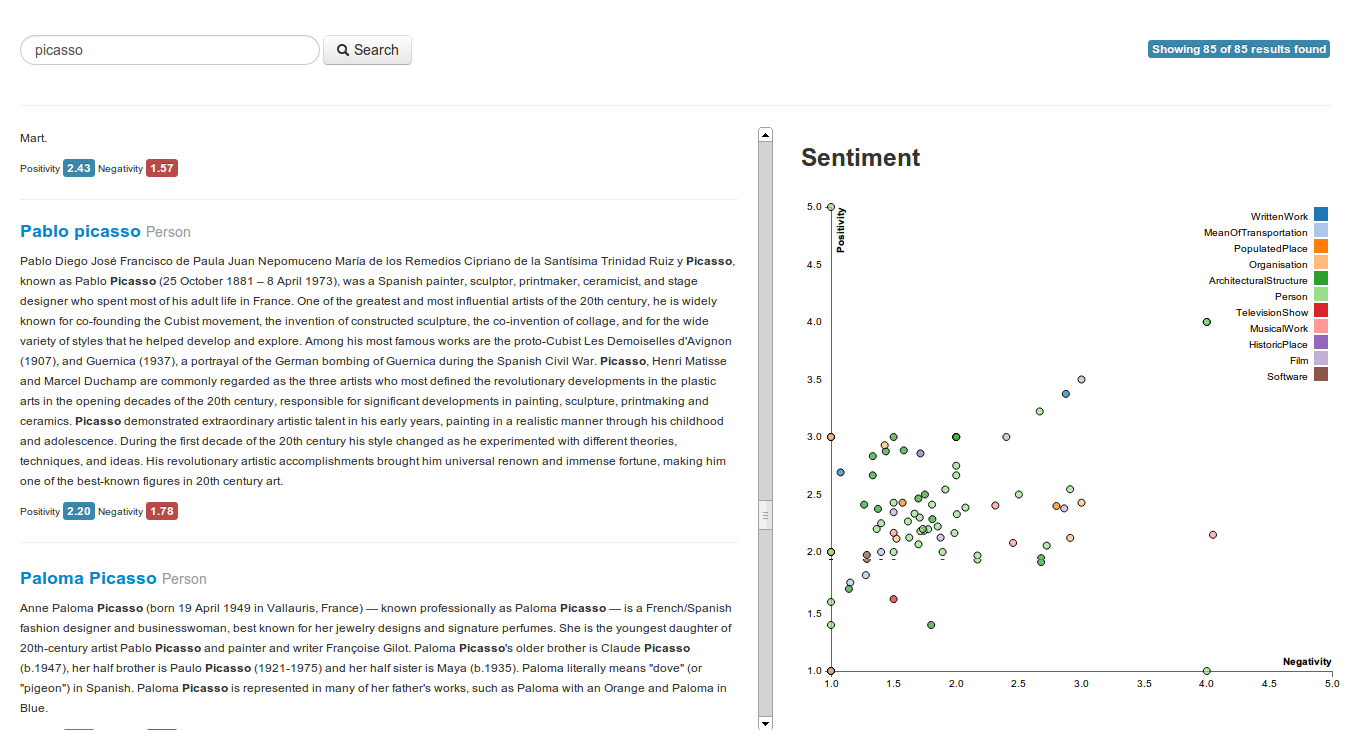}
\caption{Scatter Plot widget in our prototype search interface.}
\label{fig:search-starfield}
\end{figure}

\begin{figure}[htb]
\centering
\includegraphics[width=0.48\textwidth]{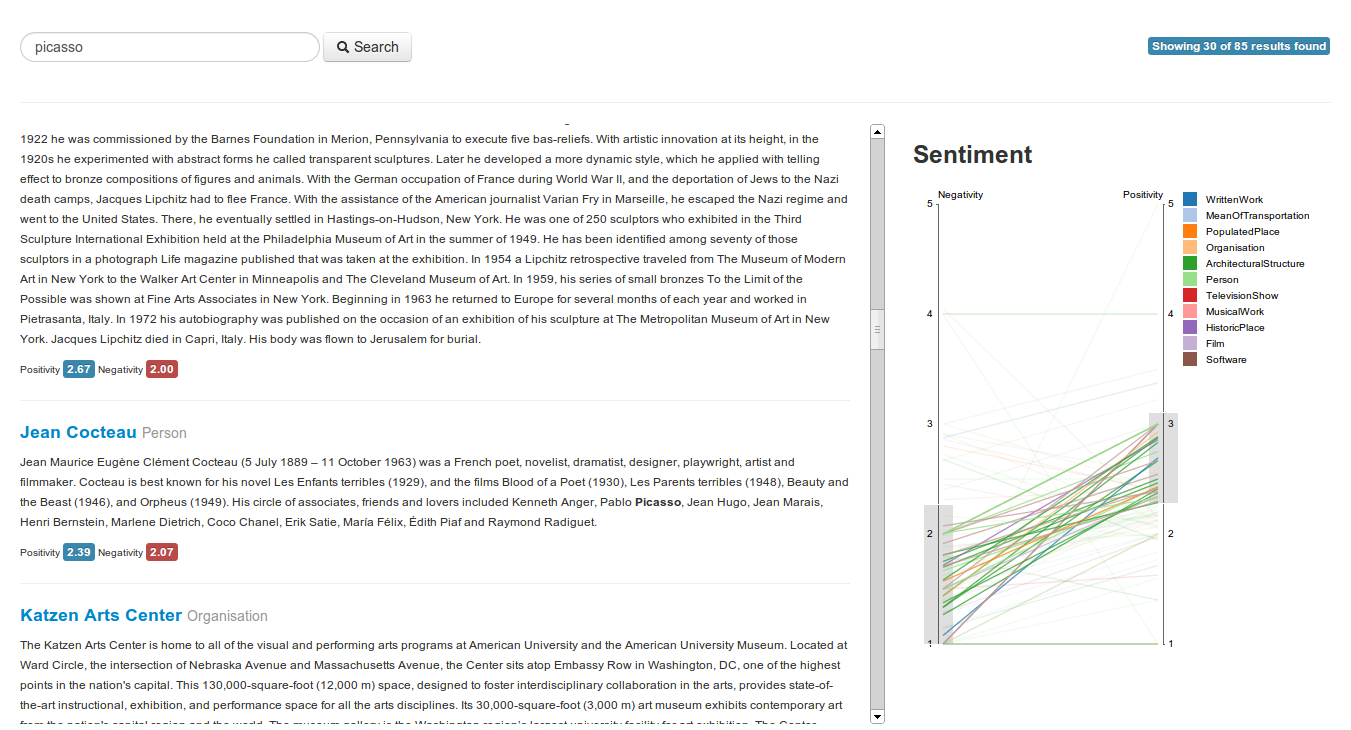}
\caption{Parallel Coordinates widget in our prototype search interface.}
\label{fig:search-parallel-coordinates}
\end{figure}

Following our design goals, we implemented two visualisation widgets:  \emph{scatter plots} and \emph{parallel coordinates} \cite{inselberg1985plane}. 
We chose two known paradigms because our research question is not about new visualisations, and using only one paradigm might bias the results of our study. 

\spara{Scatter Plot.}
Figure \ref{fig:search-starfield} shows the scatter plot widget. Each result is a circle whose position is determined by both sentiment attributes: positivity is mapped to the x-axis and negativity is mapped to the y-axis. To filter results, the information seeker can draw a rectangle over the visualisation canvas, selecting only the circles that are positioned inside the rectangle. 

\spara{Parallel Coordinates.}
Figure \ref{fig:search-parallel-coordinates} shows the parallel coordinates widget, where each attribute is a different axis: negativity is mapped to the left axis and positivity is mapped to the right axis. Each result is represented as a line that connects the corresponding value of its attributes in each axis. 
To filter results, the information seeker can draw a rectangle over the axes, selecting only the lines that begin (or end) inside the selected range. 

In both widgets we display \emph{positivity} and \emph{negativity} for each item (\emph{depict ambivalence}). 
We use transparency to showcase density and prevent occlusion (\emph{show sentiment distribution}). 
We use \emph{brushing and linking} \cite{eick1995high} to \emph{allow sentiment filtering}: when the information seeker restricts or widens the ranges of sentiment of interest in the widget, the list of results is updated immediately, and when the information seeker selects a result from the text list, the corresponding element on the visualisation is highlighted. 
The results filtered out are drawn with more transparency to indicate that they are out of focus. 
Color coding of points and lines is used to encode item categories if available.

\section{Sentiment in Wikipedia}

\begin{figure}[bhtp]
\centering
\includegraphics[width=0.45\textwidth]{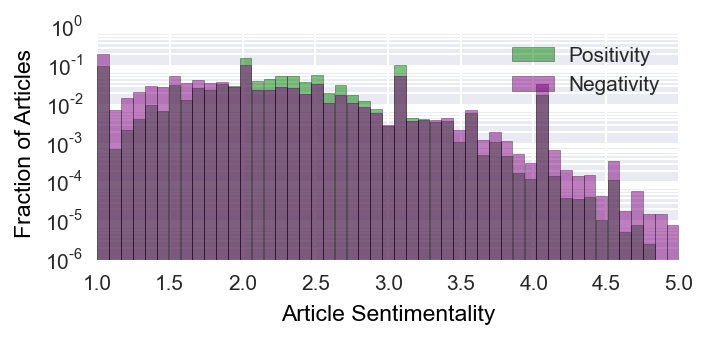}
\caption{Distributions of Positivity and Negativity in our dataset of Wikipedia articles (using log-scale).}
\label{fig:sentiment-wikipedia}
\end{figure}

We test our approach on Wikipedia\footnote{\url{http://wikipedia.org}} -- a multilingual, web-based, free-content encyclopedia, written collaboratively by a large number of volunteers.
Although Wikipedia has a \emph{neutral point of view} policy \cite{wiki:npov}, neutral is not equal to emotionless. 
It is possible to find sentiment in content in Wikipedia, as it contains biographies, disasters, awards, celebrations and summaries of fiction, among other categories. 

\spara{Dataset.}
We use a dataset of $737,863$ english articles from Wikipedia with annotated sentiment \cite{mejova2013searching}.
Each article is annotated with two scores: \emph{positivity} (from $1$ to $5$) and \emph{negativity} (from $1$ to $5$).  
Note that positivity does not imply that negativity is absent, and vice-versa: ambivalence is almost always present in text.
The sentiment values of an article are calculated based on the content of the article itself, and that of other articles linking to it.
In other words, each article becomes annotated with the sentiment scores of its own content, plus that of the associated articles.
Figure \ref{fig:sentiment-wikipedia} shows the distributions of both scores in our dataset.

Although the distributions are skewed towards lower values of sentimentality (the average positivity is $2.17 \pm 0.64$ and the average negativity is $1.92 \pm 0.78$), there are articles with high values of sentiment attributes. 
The distributions confirm that there is sentiment in Wikipedia, creating the opportunity to use our visual approach to search and explore it.

\section{User Evaluation}

We performed a small-scale usability study in a lab-setting with $13$ participants ($5$ male and $8$ female; $5$ aged $20$--$29$, $6$ aged $30$--$39$, $1$ aged $40$--$49$, and $1$ unknown), who scored their knowledge in visual web search as $3.46 \pm 1.13$ in average (using a Likert scale from $1$ to $5$). 
Participants were recruited from open calls in social networks and did not receive compensation for participating in the study. 

\spara{Apparatus.}
We built a prototype search engine that indexed extended abstracts\footnote{Defined as the first section of each Wikipedia article.} of the $737,863$ articles in the dataset. 
The user interface contained the following elements: query box, the number of results, the list of results with each article's title, extended abstract and sentiment values in text form, and the visualisation widgets. 
Given a query, the search engine returned a list of articles (maximum count: $200$) ranked using the BM25 scoring algorithm \cite{baeza2011modern}.
All participants used the same computer, a notebook of $15$ inches screen with resolution of $1440 \times 900$ pixels. 
In the experimental prototype, categorical color coding was based on the DBPedia ontology class of each article \cite{auer2007dbpedia}. This ontology is shallow, and we restricted the depth of ontologies associated to search results to be able to create a color mapping understandable for users.

\begin{figure}[thb]
\centering
\includegraphics[width=0.28\textwidth]{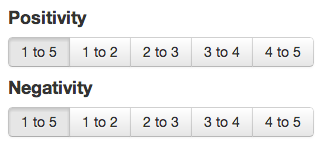}
\caption{Text-based widget used in the baseline approach of the user study.}
\label{fig:search_buttons}
\end{figure}

\spara{Design and Procedure.} 
The study used a within-subjects design. 
Each participant tested three treatments: baseline (\emph{BA}, a text-based widget of buttons to filter the results, shown in Figure \ref{fig:search_buttons}), scatter plot (\emph{SC}, shown in Figure \ref{fig:search-starfield}) and parallel coordinates (\emph{PC}, shown in Figure \ref{fig:search-parallel-coordinates}). 
The order of pairs \textit{(task, treatment)} was randomised for all participants to avoid positional bias. 

After performing each task, participants were asked to answer five questions about aesthetic value of the interface\footnote{Example: \emph{The search system was aesthetically appealing}.}. 
A Likert scale from $1$ (\emph{strongly disagree}) to $5$ (\emph{strongly agree}) was used for this purpose. 
In addition, participants were asked to write a small summary of the results they found, and were asked about their \emph{perceived time} \cite{o2009development} of task completion.
After performing all tasks, participants filled a feedback questionnaire about their thoughts on the interfaces, how they would describe the different widgets and if they had any comments and suggestions.
Finally, we logged each query and calculated the actual time of completion for each task in order to estimate the difference between perceived time and real task completion time. 
This metric is called \emph{subjective duration assessment} \cite{czerwinski2001subjective} and has been interpreted before as \emph{cognitive engagement} \cite{de2011identifying}: lesser perceived time than task completion indicates positive engagement.

\spara{Tasks.}
Participants were asked to perform three exploratory search tasks, one task per treatment. One task was personalised in terms of what they had to search for, while the remaining tasks were based on the definitions in \cite{kules2008creating}: 

\vspace{0.1cm}
{
\centering
\scriptsize
\begin{tabulary}{0.48\textwidth}{|L|}
\toprule
\emph{``Think about a topic you like, and find five articles with a highly negative connotation. Then think about a topic you do not like, and find five articles with a highly positive connotation''.} \\
\midrule
\emph{``Imagine you are taking a class called `Art in Europe'. For this class you need to
			write a research paper on some aspect of an art movement, but have yet to decide on a movement you will focus. Use the system to find three artists within that movement: one artist with a positive connotation on Wikipedia (but slightly negative), one with a negative connotation (but slightly positive), and one with high emotionality (by being highly positive and negative at the same time),  so that you might make a decision as to which movement you will write about.''} \\
\midrule
\emph{``Your professor wants you to write a paper comparing the consequences of war in three
	countries. Use the system to find three countries which have highly emotional (high positivity and negativity) events or works as consequences of the war. Find three events or works for each country.''} \\
\bottomrule
\end{tabulary}
}

\subsection{Results}

\begin{table}[t]
\center
\scriptsize
\begin{tabulary}{0.47\textwidth}{LCCCCC}
\toprule
{} &    \texttt{BA} &  \texttt{PC} &  \texttt{SC} & $K$ &  $p$ \\
\midrule
Query Count  &    $7.38$ &  $14.15$ &     $19.31$ & $5.18$ & $0.07$ \\
Task Time (s)     &  $463.00$ & $745.23$ &   $1035.92^{*}$ & $6.83$  & $0.03$ \\
Perceived Time &  $507.69$ & $770.77$ &    $761.54$ & $3.73$ & $0.15$ \\
Cognitive Engagement     &  $-44.69$ & $-25.54$ &    $274.38$ & $3.53$ & $0.17$ \\
Aesthetics & $13.54$ &  $15.77$ & $17.08$ & $2.06$ & $0.36$ \\
\bottomrule
\end{tabulary}
\caption{Experimental results for all participants ($N = 13$). $p$ values correspond to Kruskal-Wallis analysis of variance ($K$). $^{*}$: post-hoc comparison between SC and BA significant at Bonferroni corrected $p < 0.017$.}
\label{table:presential-user-study}
\end{table}

To answer our research question, \textbf{do visual approaches foster exploration in a sentiment-based exploratory search setting?}, we tested the following hypothesis: \emph{in exploration on sentiment-based scenarios, participants perform more queries and spend more time when using visualisation widgets}, by evaluating the two visualisation widgets against a text-based baseline. Post-hoc differences in means were tested using \emph{Wilcoxon's Ranked Sums} (Bonferroni corrected) after performing \textit{Kruskal-Wallis} analysis of variance on the three groups.

Results in Table \ref{table:presential-user-study} partially support our hypothesis. There is a significant group difference in task time ($p < 0.05$), and post-hoc testing revealed that SC task time is larger ($p < 0.017$, Bonferroni corrected) than the other two groups. In terms of query count, no significant effect was found, although there is a trend towards greater amount of queries in visual approaches ($p < 0.1$). Hence, using SC, users spend more time exploring, but do not necessarily perform more queries. No significant differences were found in aesthetic perception, perceived time and cognitive engagement.

\begin{table}[t]
\center
\scriptsize
\begin{tabulary}{0.48\textwidth}{LCCC}
\toprule
{} &  Achievers ($N = 6$) &  Explorers ($N = 7$) &         $p$ \\
\midrule
Total Queries     &          $25.33$ &          $54.14$ &  $0.02$ \\
Total Time       &        $1372.50$ &        $2991.26$ &  $0.00$ \\
\midrule
Queries BA        &           $6.67$ &           $8.00$ &  $0.83$ \\
Queries SC        &           $8.83$ &          $28.29$ &  $0.02$ \\
Queries PC        &           $9.83$ &          $17.86$ &  $0.13$ \\
\midrule
Task Time BA      &         $326.67$ &         $579.86$ &  $0.20$ \\
Task Time SC      &         $566.67$ &        $1438.14$ &  $0.02$ \\
Task Time PC      &         $479.17$ &         $973.29$ &  $0.03$ \\
\midrule
Perceived Time BA &         $550.00$ &         $471.43$ &  $0.78$ \\
Perceived Time SC &         $600.00$ &         $900.00$ &  $0.09$ \\
Perceived Time PC &         $620.00$ &         $900.00$ &  $0.07$ \\
\midrule
C. Engagement BA      &        $-223.33$ &         $108.43$ &  $0.02$ \\
C. Engagement SC      &         $-33.33$ &         $538.14$ &  $0.03$ \\
C. Engagement PC     &        $-140.83$ &          $73.29$ &  $0.31$ \\
\bottomrule
\end{tabulary}
\caption{Post-hoc comparison with \textit{Wilcoxon's Ranked Sums} of results for explorers and achievers.}
\label{table:explorers-achievers}
\end{table}

To explain quantitatively the differences in task time, we considered the following user taxonomy: 
\emph{achievers} (those who \emph{``are interested in doing things to the game, i.e. in ACTING on the WORLD''}), 
and \emph{explorers} (those who \emph{``are interested in having the game surprise them, i.e. in INTERACTING with the WORLD''}) \cite{bartle1996hearts}. 
We define achievers ($N = 6$) as those users who are in the bottom $50\%$ w.r.t. the geometric mean of total task time and total queries issued; and explorers as the rest ($N = 7$), that is, those in the upper $50\%$.
In this way, achievers want to finish the task fast and quickly, while explorers are interested to see how the system can surprise them.
Table \ref{table:explorers-achievers} reports differences in means for both groups in all approaches. 
There are significant differences (measured with \textit{Wilcoxon's Ranked Sums}) on total queries and total time, which were expected as they are consequence of the user taxonomy.
However, other significant differences emerge: 
\emph{1)} explorers issue more queries than achievers using SC ($p < 0.05$), but not in BA and PC;
\emph{2)} explorers spend more time when using SC and PC ($p < 0.05$) but not when using the baseline;
\emph{3)} explorers have greater positive cognitive engagement than achievers when using SC and BA (and SC's engagement is almost $5$ times BA).

\spara{Qualitative Feedback.}
We included open-feedback questions in order to understand and explain the quantitative results. We use [\texttt{P$i$}] to refer to participant $i$.

The baseline (BA) was characterised by participants as \emph{``boring''} [\texttt{P8}] but \emph{``the easiest for me to find results''} [\texttt{P4}]. It was perceived as a tool for \emph{``discriminating''} [\texttt{P11}] and \emph{``filtering''} [\texttt{P10}].
As expected, the \emph{``filters were really easy to use''} [\texttt{P3}], as participants are used to this kind of interface. 
However, most of the positive feedback for BA was related to the act of performing the task, and not on how the actual users felt about the text-based widget: \emph{``I think the most useful one is the buttons one because it has more precise information reflected on it.''} [\texttt{P10}], although not everyone felt comfortable with it: \emph{``The one with the numbers was misleading for me''} [\texttt{P6}].

Regarding the visual approaches, the scatter plot (SC) was described as \emph{``attractive''} [\texttt{P8}], \emph{``like a classifier''} [\texttt{P4}], as well as a \emph{``spectrum''} [\texttt{P10}] or a \emph{``map''} [\texttt{P11}], perhaps referring to how a scatter plot allows to classify elements according to their position on the screen.
We expected that users would have been familiar with SC, as in: \emph{``[BA] and [SC] are easy to use. They are helpful and easy to understand''} [\texttt{P3}]. However, it also \emph{``needs more concentration''} [\texttt{P6}]. Some users were more vocal in their enthusiasm for this approach: \emph{``this is the task that I enjoy the most! I liked pretty much the graphics''} [\texttt{P8}], \emph{``this is the approach I liked the most, it was easier to filter the results''} [\texttt{P9}], indicating that scatter plots not only are familiar, but also they generate a more positive, emotional reaction.
Parallel Coordinates (PC) produced an ambivalent reaction. On one hand, it was described as \emph{``interesting''} [\texttt{P8}],\emph{``much more cooler then the other one''} [\texttt{P1}], \emph{``the high-low thing helps me to know if it is positive or negative faster. I really like how [PC] worked''} [\texttt{P6}], and \emph{``the sentiment indicator [PC] helps in the task''} [\texttt{P9}]. On the other hand, users claimed that \emph{``[PC] was not appealing nor easy to understand or use''} [\texttt{P3}] and that \emph{``it's confusing''} [\texttt{P11}].

In addition to visualisation feedback, participants suggested some features that could improve our system prototype: 
\emph{``drawing the box around the numbers in each axis is too complex, I would have preferred to have another way of controlling the sentiment in the results. Maybe even a simple slider''} [\texttt{P9}, referring to PC], \emph{``a grid in the circles system would help to have more exact information about the scores at a glance''} [\texttt{P11}, referring to SC]. 
With respect to the search results, some users expressed they were not satisfied with their quality: 
\emph{``the search engine does not work properly, distracting myself from the task''} [\texttt{P3}],
\emph{``the search was very frustrating, as the searches often did not yield many results''} [\texttt{P7}].
Some users thought about whether they would use a system like this in the future:
\emph{``the system was useful but I don't search using sentiments frequently\ldots Maybe when searching for the politic situation of a country I would use it''} [\texttt{P9}].

\section{Discussion and Implications}
\spara{User Engagement and Visualisation Widgets.}
In the experiment the SC group spent more time performing the exploratory tasks than BA and PC. 
Whether this is a good scenario, if users spend more time because they are engaged, or if they spend more time because the visualisation is impeding the task at hand, is something that needs to be determined and explained.
We attributed part of the longer task time of SC in Table \ref{table:presential-user-study} to a positive user experience \emph{when users are explorers}, as explorers performed more queries and spent more time, while at the same time they showed a significantly greater positive cognitive engagement.
Moreover, the qualitative feedback received by SC was positive, indicating that it is unlikely a negative experience when using that treatment to perform the task.
This positive engagement result is consistent with previous work \cite{clarkson2009resultmaps}, where users expressed more enjoyment when using visualisation techniques in the search interface.
Since not all visualisations are perceived equally, it makes sense that some visualisations engage users and some do not, as well that a visualisation might engage one kind of users only.
In this aspect, our results are limited to explorers only, because no significant patterns were found for achievers, although some users explicitly favored the parallel coordinates widget as attractive. 
As there might not be a globally \emph{better} visualisation for all users, it remains to be seen which visualisation is more likely to engage \textit{achievers}.

\spara{Personalisation of User Interfaces.}
Individual differences based on exploratory behavior provide a base for a contextual personalisation of user interfaces, as a to complement to content personalisation based on user generated content.
When considering individual differences, we restricted the definition of exploration as the geometric mean of task time and number of queries, which allows to implement the \textit{explorers and achievers} taxonomy based on:
\emph{1)} \textit{previous activity on the search system}, making possible to provide this type of widget-based personalisation when query logs and interaction data are available;
\emph{2)} \textit{granularity of a query}, in the sense of how \emph{``good latin restaurant in Born neighborhood''} indicates something one wants to achieve, while \emph{``restaurants in Barcelona''} indicates something one wants to explore.
Considering availability of this user taxonomy, user interfaces can be personalised to increase engagement in users performing learning and investigation tasks by using scatter plots instead of text widgets.

\spara{Limitations.}
In terms of implementation, participants in our experiment expected better results than those provided by our prototype implementation.
The effect of those unfulfilled expectations over the obtained results is unknown and should be considered in future experiments.
In addition, trending differences in behavior surfaced on quantitative results, perhaps a limitation of the small-scale of the user study. 
We believe these limitations can be fully addressed in a larger-scale experiment using an improved search engine and following the TREC interaction track guidelines \cite{dumais2005trec}.

\section{Conclusions}
This paper presented results on our research of sentiment visualisation widgets for exploratory search. 
We defined design goals in this scenario, and implemented two visualisations based on known techniques: scatter plots and parallel coordinates.
Both approaches were evaluated against a baseline of text-based links for exploring search results. 
Even though the scale of our study is small, we found statistical evidence of users spending more time performing tasks when using scatter plots.
Through analysis of qualitative feedback and individual differences, we explained that time difference as positive engagement with the visualisation widget.
In particular, the individual differences analysis focused on a user taxonomy that defines \emph{explorers} and \emph{achievers}: those who \emph{interact} in the world and those who \emph{act} in it, respectively.
Our results indicate that scatter plots are suitable for explorers, as they are more engaged in a positive way when using that visualisation paradigm in comparison to a text baseline and the parallel coordinates visualisation.
Hence, in the presence of explorers, we suggest search and exploratory systems to personalise the user interface with scatter plots to browse sentiment, to increase user engagement and foster exploration.

\spara{Future Work.}
Our approach assumes the presence of sentiment meta-data, which may be added algorithmically to any text collection. 
The usage of Wikipedia proves to be useful as there is a varying degree of sentimentality across the subset we studied. 
As future work we will consider other scenarios, such as reviews, media and social networks, where the amount and variation of sentiment will likely be greater.
In addition, we will consider more complex behavioral taxonomies based on personality traits \cite{heinstrom2002fast}, as personality traits in social networks can be predicted in social media \cite{quercia2011our,quercia2012personality}.
Finally, we will explore the possibilities of our approach in other bivariate related contexts such as political leaning.

{
\spara{Acknowledgments.} 
This work was partially funded by Grant TIN2012-38741 (Understanding Social Media: An Integrated Data Mining Approach) of the Ministry of Economy and Competitiveness of Spain.
}

{
\small
\bibliographystyle{plain}
\bibliography{sentiment_vis}
}
\end{document}